\documentclass[twocolumn]{IEEEtran}

\usepackage{cite}
\usepackage{graphicx}
\usepackage{amsmath,amssymb,amsthm}
\usepackage{multirow}

\usepackage[rgb]{xcolor}

\begin{document}

\title{What Role Do Intelligent Reflecting Surfaces Play in Multi-Antenna Non-Orthogonal Multiple Access?}


\author{
Arthur S. de Sena, \textit{Member}, \textit{IEEE}, Dick Carrillo, \textit{Member}, \textit{IEEE}, Fang Fang,  \textit{Member}, \textit{IEEE}, Pedro H. J. Nardelli, \textit{Senior Member}, \textit{IEEE}, Daniel B. da Costa, \textit{Senior Member}, \textit{IEEE},
Ugo S. Dias, \textit{Senior Member}, \textit{IEEE}, Zhiguo Ding, \textit{Fellow}, \textit{IEEE}, Constantinos B. Papadias, \textit{Fellow}, \textit{IEEE}, \\and Walid Saad, \textit{Fellow}, \textit{IEEE}

\thanks{A. S. de Sena, Dick Carrillo, and P. H. J. Nardelli are with the Lappeenranta-Lahti University of Technology, Finland.}

\thanks{F. Fang is with Durham University, UK.}

\thanks{D. B. da Costa is with the Federal University of Cear\'{a}, Brazil.}

\thanks{U. S. Dias is with the University of Bras\'{i}lia, Brazil.}

\thanks{Z. Ding is with the University of Manchester, UK.}

\thanks{C. B. Papadias is with the American College of Greece, Greece.}

\thanks{Walid Saad is with Virginia Tech, USA.}
}

\maketitle

\begin{abstract}
Massive multiple-input multiple-output (MIMO) and non-orthogonal multiple access (NOMA) are two key techniques for enabling massive connectivity in future wireless networks. A massive MIMO-NOMA system can deliver remarkable spectral improvements and low communication latency. Nevertheless, the uncontrollable stochastic behavior of the wireless channels can still degrade its performance. In this context, intelligent reflecting surface (IRS) has arisen as a promising technology for smartly overcoming the harmful effects of the wireless environment. The disruptive IRS concept of controlling the propagation channels via software can provide attractive performance gains to the communication networks, including higher data rates, improved user fairness, and, possibly, higher energy efficiency. In this article, in contrast to the existing literature, we demonstrate the main roles of IRSs in MIMO-NOMA systems. Specifically, we identify and perform a comprehensive discussion of the main performance gains that can be achieved in IRS-assisted massive MIMO-NOMA (IRS-NOMA) networks. We outline exciting futuristic use case scenarios for IRS-NOMA and expose the main related challenges and future research directions. Furthermore, throughout the article, we support our in-depth discussions with representative numerical results.
\end{abstract}

\section{Introduction}

The fifth generation (5G) of wireless cellular systems will enable the deployment of demanding applications such as autonomous cars, massive sensor networks, telemedicine, smart homes, and more. To make these applications possible, stringent requirements such as massive connectivity, improved spectrum efficiency, and low communication latency must be fulfilled. Massive multiple-input multiple-output (MIMO) is one of 5G's key technologies for accomplishing these requirements. By exploiting the spatial domain with transmit beamforming techniques, and employing a large number of antennas, massive MIMO schemes enable resource-efficient parallel transmissions to multiple users using the same frequency and time-slot. Non-orthogonal multiple access (NOMA) is another important technology envisioned to be part of future wireless systems. In particular, by employing superposition coding (SC) at the base station (BS) and successive interference cancellation (SIC) at the receivers, power-domain NOMA can simultaneously serve more than one user with a single resource block. This makes NOMA capable of providing significant connectivity improvements to the communication networks. If massive MIMO and NOMA are properly combined, the features of the two techniques can be exploited to reach even higher spectral gains, which can outperform conventional systems employing orthogonal multiple access (OMA) \cite{ref1}.

Nevertheless, despite their potential improvements, a \emph{MIMO-NOMA} system has several limitations. The random fluctuation of wireless channels, signal path loss, high user mobility, and atmospheric absorption are just a few examples of issues that can strongly impact the performance of MIMO-NOMA systems \cite{ref2}. The impact of such impairments becomes even more severe at higher frequencies, i.e., frequencies above $6$ GHz, which are a key feature of 5G systems and beyond. Although an increase in the number of antennas can help to overcome such kinds of performance degradation, this comes at the cost of an increased energy consumption when the number of antenna elements becomes high. In view of this, while 5G is still being deployed, engineers and researchers have already started looking at new energy-efficient technologies to go beyond 5G and build the sixth generation (6G) \cite{ref10}. In particular, due to recent advances in the field of electromagnetic metamaterials, the appealing concept of an intelligent reflecting surface (IRS) has been drawing significant attention from both academia and industry \cite{ref5}.

{
\begin{figure*}[t]
	\centering
	\includegraphics[width=1\textwidth]{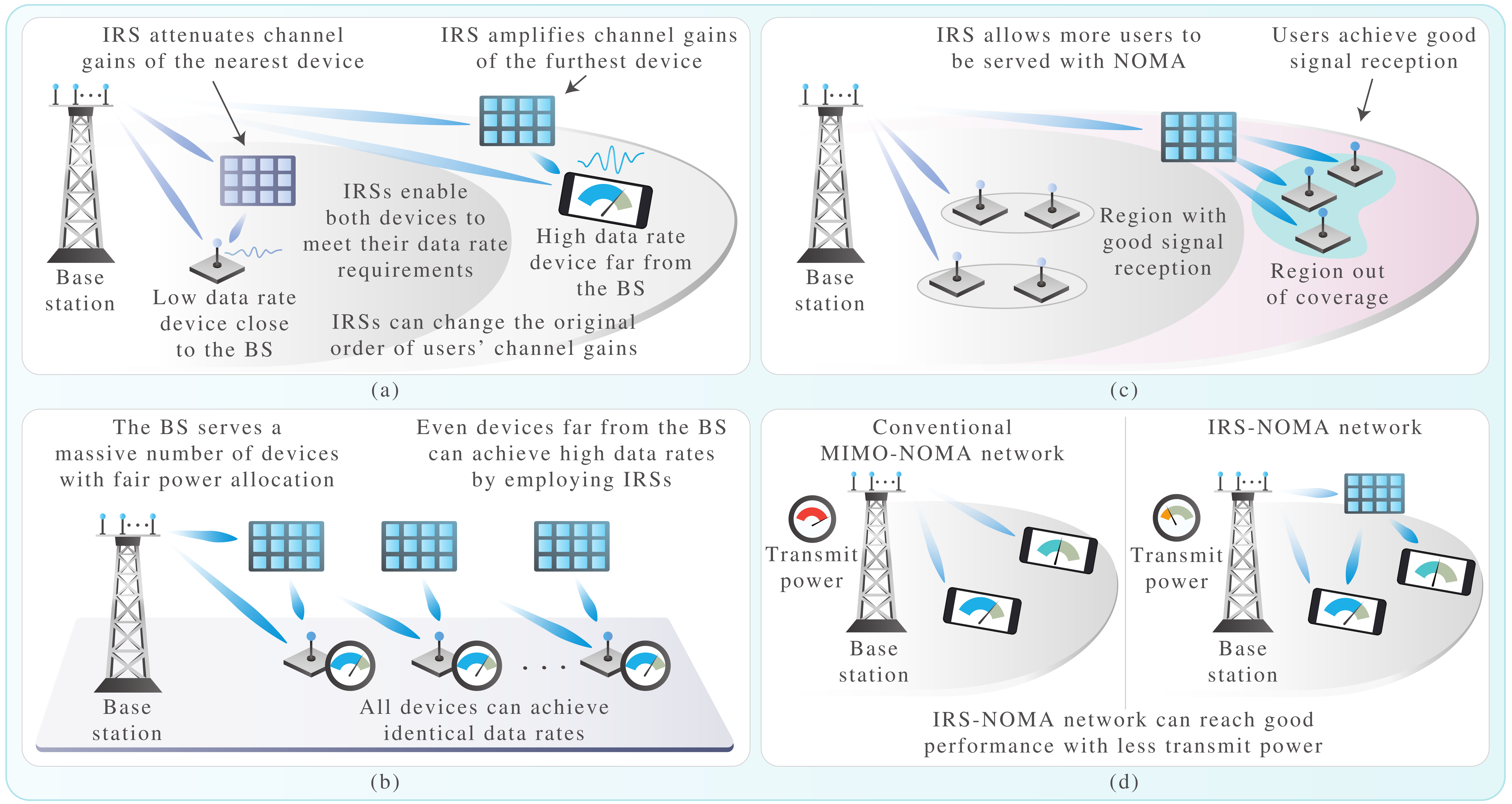}
	\caption{Illustration of potential achievements of IRS-NOMA networks. (a) The IRSs can tune the channel gains to meet the data rate requirements of both near and far users. (b) The IRSs can boost the performance of the network under fair resource allocation. (c) By deploying an IRS at the cell-edge, more users can be served with NOMA. (d) The IRS-NOMA system can achieve similar rate performance of conventional MIMO-NOMA with less transmit power.}\label{achiev}
\end{figure*}
}

An IRS is an ultra-thin planar structure composed of a large number of reflecting elements with a size smaller than the signal wavelength, known as meta-atoms \cite{ref4}. The key advantage of IRS structures is that each meta-atom can be dynamically tuned by software with distinct phases and amplitudes of reflection so that they can collaboratively forward the impinging waves with, ideally, any desired radiation pattern, like a hologram. Such a capability enables the deployment of smart wireless environments with optimized and possibly energy-efficient signal propagation, thus paving the way towards a new wireless communication paradigm. For instance, by properly optimizing the IRS's reflection coefficients, signal beams can be formed to achieve goals such as to enhance the performance of a specific terminal, to cancel interference coming from other devices, or even to completely null out information leakage at an eavesdropper \cite{ref7,ref11}. In addition to these advantageous features, an IRS has potential to exhibit near-zero energy consumption since it does not comprise the power-hungry components of conventional radiofrequency chains. In general, an IRS design contains passive metamaterial parts and an active control components with only ultra-low-power electronic circuitry that can be powered by energy harvesting wireless modules \cite{ref5}. These make IRSs a promising plug-and-play technology for improving the performance of future communication networks.

Although a number of surveys on the topic of IRS have recently appeared in \cite{ref5}, \cite{ref7}, and \cite{refnew1}, and a few technical works on IRS and MIMO-NOMA  exist \cite{ref4,ref9}, there are no tutorials or surveys that overview and study these two subjects combined. As a result, it is still not completely clear what are the roles that IRSs can play in MIMO-NOMA networks. Therefore, further studies for the in-depth understanding of the combination of these two promising technologies are required. In view of this, the main contribution of this article is to investigate the potential spectral and energy efficiency gains of IRS-assisted massive MIMO-NOMA (\emph{IRS-NOMA}) systems in future wireless networks as presented in Fig. \ref{achiev}. We will also discuss the fundamental challenges pertaining to their effective deployment. In summary, this article will have the following key contributions:

\begin{itemize}
\item Four attractive potential achievements of IRS-NOMA networks are identified. Specifically, we show that IRS-NOMA can enable flexible control on the users' channel gains, enhanced user fairness, enhanced scalability, and improved energy efficiency. 

\item  Pervasive coverage via multiple IRSs, 3D coverage in UAVs networks, and massive grant-free transmissions are introduced as  three promising applications that IRS-NOMA has potential to enable   in future wireless systems beyond 5G.

\item A comprehensive discussion of the main open problems, technical challenges, and possible future directions of IRS-NOMA is provided.
   
\end{itemize}

	\begin{table*}[th]
	\caption{Comparison of IRS with other technologies}
	\label{IRSVsRelay}
	\renewcommand{\arraystretch}{1}
	\centering
	\begin{tabular}{|c|c|c|c|}
		\hline 
		\multirow{2}{2.2cm}{\textbf{Technology} }& \multirow{2}{*}{\textbf{Operation mode}} & \multirow{2}{2.2cm}{\textbf{Characters}}&\multirow{2}{*}{\textbf{Drawbacks}}\\ 
		& & &\\
		\hline	\multirow{2}{*}{IRS} & \multirow{2}{*}{Full duplex} & Low hardware cost  &  Short range of implementation\\ 
		& &Potential to exhibit low energy consumption &Difficult to estimate CSI\\
		\hline
		\multirow{2}{2.2cm}{AF relay} & \multirow{2}{2.2cm}{Half/full duplex}  & \multirow{2}{*}{Actively regenerate and transmit signals}  & {High hardware cost} \\
		& & &High energy consumption\\
		\hline
		\multirow{2}{*}{Ambient Backscatter} 
		& \multirow{2}{2.2cm}{Half duplex}&Low hardware cost & Limited data rate \\ 
		& & Low energy consumption& Strong interference from active source\\
		\hline		
	\end{tabular}
\end{table*}

\section{An overview on the IRS Technology}
To shed light on the operation of an IRS system, in this section, we provide a fundamental background on IRS hardware architecture, channel model, and a comparison with two different related technologies. 

\subsection{Fundamentals of IRS's Architecture}
The IRS architecture is not a unified subject. The available literature proposes a variety of IRS designs with a different number of layers and different technologies, including liquid crystals, microelectromechanical systems, doped semiconductors, and electromechanical switches \cite{ref5}. Despite that, the majority of the architectures share at least three common layers, which consist of (1) a meta-atom layer, comprising a larger number of passive conductor elements and low power active switches; (2) a control layer, which is responsible for adjusting the amplitude and phase shift of each meta-atom element; and (3) a gateway layer (or communication layer) that establishes the communication between the control layer and the BS. Each meta-atom acts as a sub-wavelength scatterer with reconfigurable electromagnetic properties. Such a feature enables them to collectively change the induced current patterns in the IRS so that a desired electromagnetic field response can be generated. This allows the IRS to manipulate the wavefronts to achieve objectives like steering, absorption, polarization, filtering, and collimation \cite{refnew1}.

\subsection{IRS's Channel Model,  Estimation Strategies, and \\Associated Tradeoffs}\label{iiib}
Propagation channels via an IRS behave differently from those observed in conventional communication systems. First, in a classical cellular network, users are usually considered to be located far from the transmit antennas. As a result, most of the conventional channel models rely on the assumption that the system is operating in the far-field regime, i.e., the impinging signals at the receive antennas are approximated as plane wavefronts. On the other hand, in IRS-assisted communication systems, one cannot guarantee that users are positioned far enough from an IRS, and the far-field assumption may not always hold. More specifically, depending on the distance between the user and its serving IRS, as well as on the size of the IRS, the system might operate instead in the near-field radiative regime. The mutual coupling effect among IRS elements is another important effect that differentiates IRS systems from classical counterparts. All of these characteristics make classical multi-path and path loss channel models not suitable and inaccurate for modeling IRS-assisted communications. Therefore, the development of more realistic models is necessary to capture the fundamental performance limits of IRS systems in practice. This is in fact known to be an important open problem in this area.

In a general IRS-aided MIMO design, the channel matrix from the BS to the user, via an IRS, conventionally includes the channel responses from BS to the IRS, a diagonal matrix that models the IRS's signal reflection, and the channel responses between the IRS and the user. Specifically, the IRS receives the signal from the BS, and then reflect the impinging signal by inducing the amplitude and phase changes adjusted by the control layer. As a result, the BS-IRS-user link can be represented by a multiplicative channel model, which can be added coherently with the direct link from the BS to either boost or attenuate the signal strength at the receiver \cite{ref7}.
For a practical channel model, propagation phenomena as the discussed above, i.e., far-field effect, mutual coupling, path loss, and multi-path, should be all incorporated in these channel matrices.

To enable the real-time capabilities of an IRS, its control layer needs to optimize the meta-atom elements based on the channel state information (CSI) of the entire system, including all propagation links. To obtain such global CSI, different strategies with different tradeoffs can be employed. For instance, it is possible to estimate the BS-IRS-user link directly on the IRSs, allowing them to reconfigure their elements autonomously. To accomplish this feature, each IRS needs to comprise at least low-power sensors and must have some processing capabilities. This distributed design can facilitate, to some extent, the channel estimation process if compared to other strategies. However, the IRS optimization complexity increases as the number of elements becomes high, which can result in high energy consumption at the IRSs. Moreover, hardware complexity and costs will also increase. Equipping the BS with a central controller and employing sophisticated channel estimation protocols is another common approach used to optimize the IRSs. In this centralized design, the estimation and optimization protocols are executed at the BS, which can afford high computational power. Once the estimation and optimization are completed, the BS only needs to send the result with the optimal set of coefficients to the IRS's control layer. The main advantage of a centralized design is that, without the need for sensing components, the IRS hardware can be further simplified, which can potentially lead to lower energy consumption. The downside is that, when the number of IRS elements grows, the channel estimation complexity and the signaling overhead from the BS to the IRSs will increase, which can be challenging in practical implementations.

\subsection{IRS and Related Technologies}
Next, we discuss some important features that make IRSs distinct from related technologies like amplify-and-forward (AF) relaying and ambient backscatter radio systems.

In AF relaying networks, when a relay node amplifies the received signals (which can be energy-consuming), it also amplifies noise, which consequently can degrade the system performance. Moreover, AF relays can only operate in full-duplex mode if efficient self-interference cancellation techniques are employed. This impairment can increase the implementation cost and system complexity \cite{ref7}. In contrast, the IRSs operate in passive reflecting mode and do not require a dedicated energy source to retransmit the impinging signals. Such characteristic enables the IRSs to work in a full-duplex mode without generating self-interference and noise amplification. Ambient backscatter communication \cite{RFID2009} is another technology that operates recycling the impinging electromagnetic waves. However, the working principle and objectives of such systems are very distinct from the IRS concept. While IRSs are designed only to reflect transmitted signals, the passive backscatter devices harvest energy from the received analog waves coming from different active sources to transmit its own information. Furthermore, backscatter systems are susceptible to strong direct interference generated from the active sources, an issue that is not present in IRS networks. A comparison of the features of IRS and the aforementioned technologies is summarized in Table~\ref{IRSVsRelay}.

\section{IRS-NOMA Networks: Potential Improvements}\label{secpi}
In this section, we raise four important performance gains that IRS-NOMA systems can potentially provide, namely tuned channel gains, improved fair resource allocation, enhanced coverage range, and high energy efficiency. These attractive achievements are illustrated in Fig. \ref{achiev}. For each of the illustrated gains, we perform a comprehensive and in-depth discussion that is supported by numerical results generated from IRS-NOMA Monte Carlo simulations. In particular, we consider the downlink transmissions of a cellular network having a single BS equipped with a uniform linear array of $80$ transmit antennas that serves users with $4$ receive antennas. Unless stated otherwise, an IRS with $20$ reflecting elements is installed nearby to each user.
Given that an IRS can forward the impinging signals with high directivity and considering that users are separated far enough, we assume
that the IRS of a given user does not interfere with other IRSs.
Furthermore, we present results considering both fixed and optimized reflection coefficients. In the optimized IRS results, depending on the system objectives, we dynamically tune the reflecting elements to either maximize or attenuate the instantaneous rates achieved by each user, and, on the fixed setups, all reflection phases and amplitudes are adjusted to $0^\circ$ and 1, respectively.
To realize these capabilities, we assume a centralized architecture where the IRSs are coordinated by a central controller installed at the BS, which can accurately estimate the CSI of all propagation links.
Moreover, analogously to \cite{ref1} and \cite{ref2}, the multi-antenna users are considered to be distributed among multiple clusters, in which the users are sub-divided into multiple NOMA groups. In order to cancel inter-cluster interference, the BS employs a precoder that is constructed based on the null-space spanned by the channel matrices of interfering clusters, and each user adopts a zero-forcing receiver to eliminate the remaining inter-group interference. More details about this transmission and reception strategy can be found in \cite{ref1}. 

\subsection{Tuned Channel Gains}
In order for NOMA to be effective, during the SC process at the BS, the users are sorted in an ascending or descending order based on their channel gains so that, relying on this information, the users can successfully employ SIC to recover the transmitted messages. This renders the performance of MIMO-NOMA systems highly dependent on the users' channel conditions. Specifically, it has been shown that NOMA can achieve higher spectral efficiency than OMA only if the channel gains of different users are significantly distinct and if their spatial directions are not orthogonal to each other \cite{ref9}. The challenge of these constraints in conventional networks is that they cannot always be satisfied. This is because the highly stochastic propagation paths are determined almost exclusively by the scattering environment and the location of the receivers, in which classical communication systems have no control. This scenario completely changes when it comes to IRS-assisted networks. By deploying IRSs, the propagation environment can be smartly tuned according to the desired objectives, potentially enabling the network to finely adjust its users' channel gains so that NOMA can always achieve good spectral efficiency. For instance, the recent work in \cite{ref9} has shown that, with the help of IRSs, it is possible to force the channels to become quasi-degraded, which is a condition in which MIMO-NOMA can achieve the same performance of dirty paper coding (DPC), i.e., where it can approach the capacity region of the downlink channel.

\begin{figure}[t]
	\centering
	\includegraphics[width=.5\textwidth]{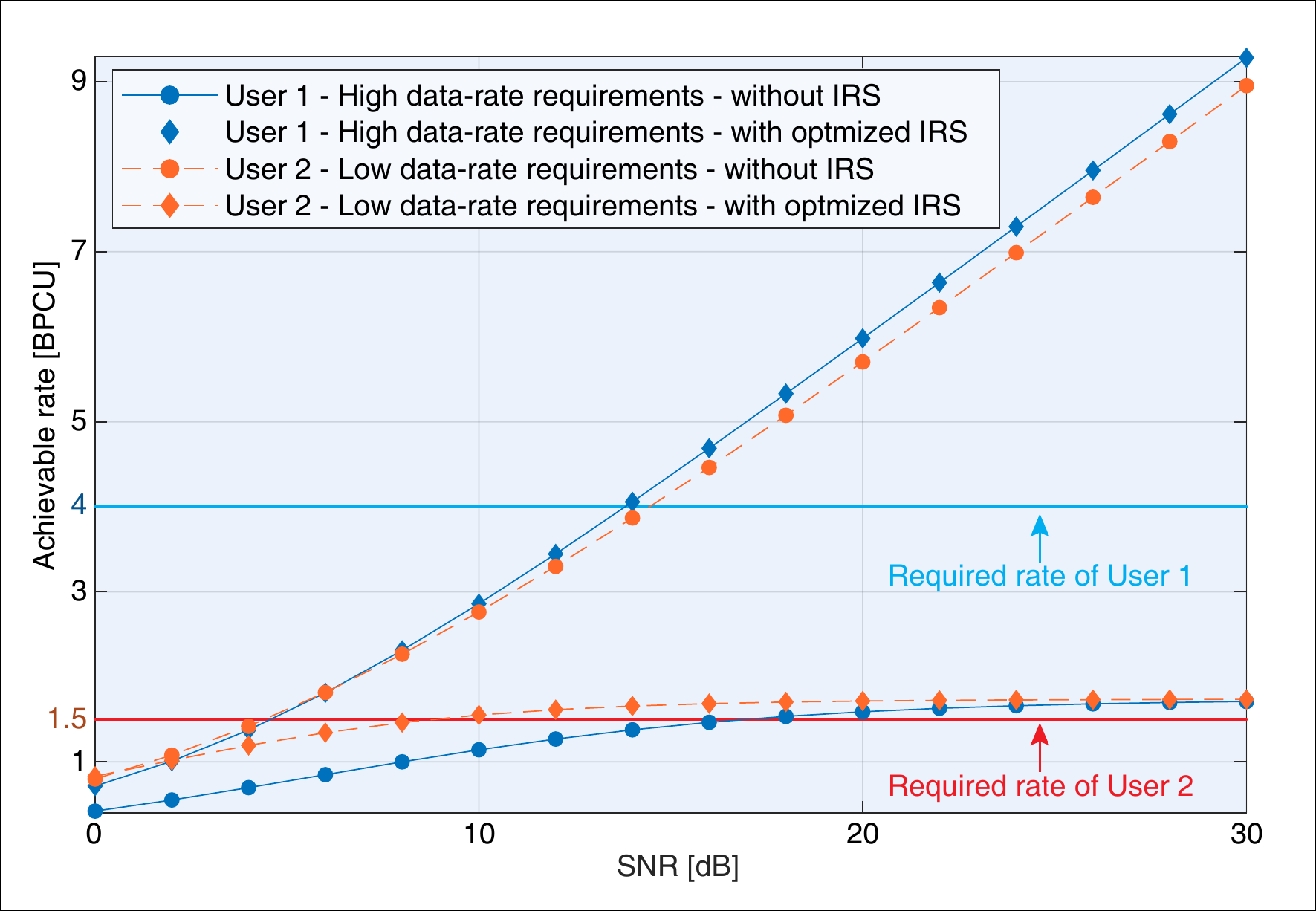}
	\caption{Achievable rates versus SNR for MIMO-NOMA and IRS-NOMA systems under fixed power allocation where user 1 is located at $200$~m and user 2 at $100$~m from the BS. The power allocation coefficients of users $1$ and $2$ are $7/10$ and $3/10$, respectively.}\label{g1}
\end{figure}

The IRS technology introduces a new paradigm to MIMO-NOMA networks by providing it with flexibility in multiplexing users. By employing IRSs, it can become possible even to change the original order of the users' channel gains. This capability enables the network to sort its users based on their particular data rate requirements rather than on the uncontrollable random environment of classical communication systems. For instance, in conventional NOMA deployments with fixed power allocation, when a user with high capacity requirements faces highly unfavorable channel conditions, e.g., when the user is very far from the BS, it will inevitably fall in an outage state (a state where its minimum performance requirements are not met). This is a difficult situation to be solved by traditional approaches since, independently of the power allocated, the weak user's rate will always be limited due to interference from the strong user. Fig. \ref{g1} shows the simulation results for the scenario illustrated in Fig. \ref{achiev}a, where user $1$ is located at $200$~m and user $2$ at $100$~m from the BS. We see that the conventional MIMO-NOMA system is not able to deliver the required rate of 4 bits per channel use (BPCU) for user $1$. By employing IRSs in this scenario, the system performance can be efficiently optimized. As one can notice, by properly adjusting the reflection coefficients of the IRSs, user $1$ that was originally experiencing bad channel conditions can achieve high performance and meet its required data rate. In contrast, the channel gains of the nearer user $2$ are optimized to provide just the necessary capacity.

\subsection{Improved Fair Power Allocation}
In some emerging applications, such as the Industrial Internet of Things, it can be important that all devices experience similar data rates. It has been demonstrated that, by properly performing power allocation, MIMO-NOMA systems can achieve this interesting capability. Specifically, the performance of different devices can be balanced by maximizing the minimum achievable rate in the network so that everyone can experience similar rate levels. The main disadvantage of such approaches is that to increase the performance of a weak device, the ones with good channel conditions can be excessively penalized. In addition, if the channel conditions of the weak device are too degraded, the data rates achieved under fair power allocation can be not enough to meet the quality of service requirements of other devices, leading to poor network performance. As illustrated in Fig. \ref{achiev}b, installing IRSs in such deployments can be very beneficial. Theoretically, all devices could reach the same data rate with the help of dynamic fair power allocation, while the IRSs could boost the channel gains to guarantee a high network performance.
	
\begin{figure}[t]
	\centering
	\includegraphics[width=.5\textwidth]{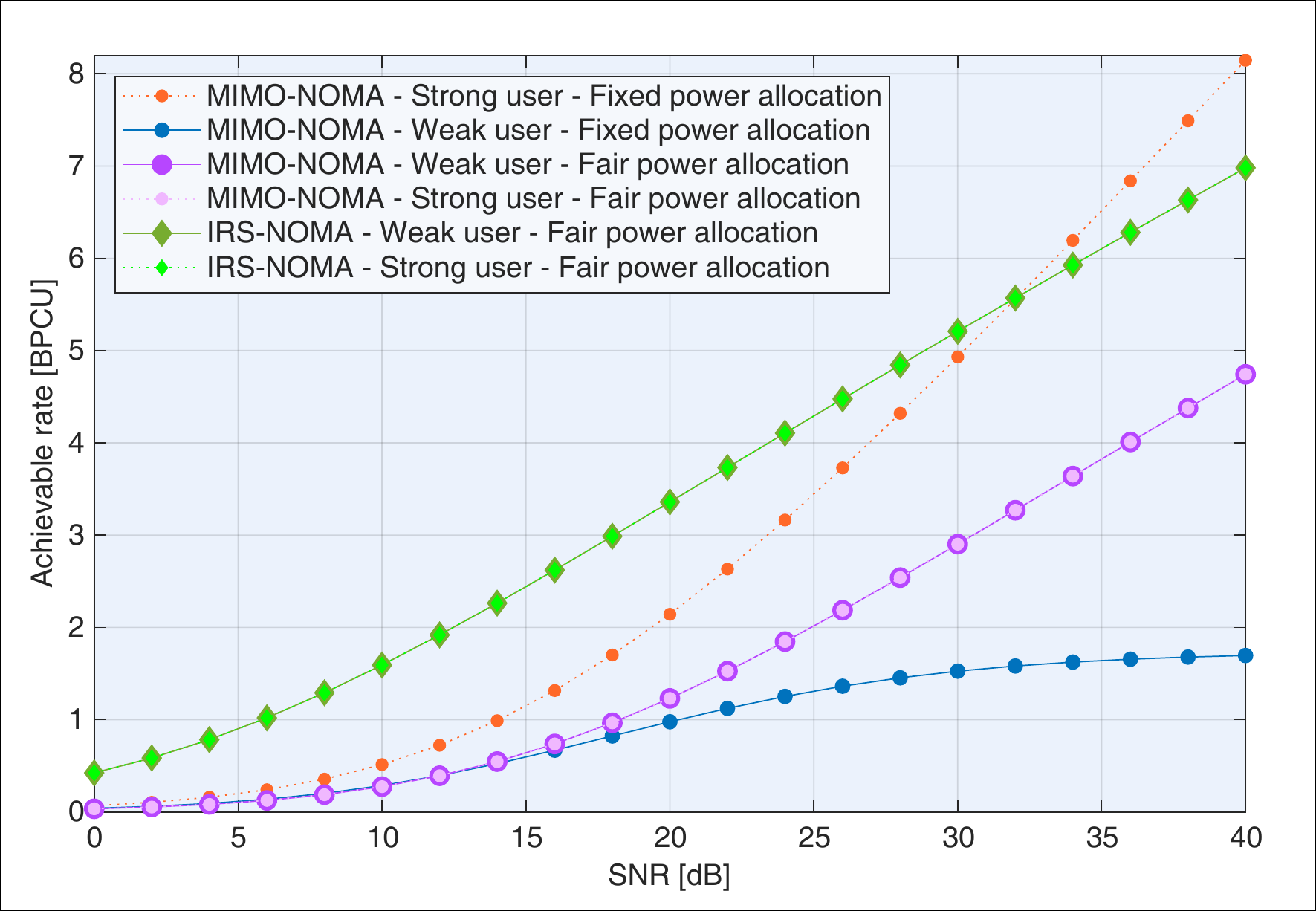}
	\caption{Achievable rates versus SNR for MIMO-NOMA and IRS-NOMA systems under fixed and fair power allocation, where the weak user is located at $200$~m and the strong one at $100$~m from the BS. When fixed allocation is employed, the power coefficients for the weak and strong users are $7/10$ and $3/10$, respectively.}\label{g2}
\end{figure}

This above achievement can be visualized in the simulation results presented in Fig. \ref{g2}, where we employ to MIMO-NOMA and IRS-NOMA both fixed and the fair power allocation policy developed in \cite{ref12}. Here, we consider the existence of two users per NOMA group, one located at $100$~m and another at $200$~m from the BS. One can see that the fair power allocation in the MIMO-NOMA system can successfully balance the users' rates. However, the strong user pays an expensive price for enabling this capability. For instance, when the signal-to-noise ratio (SNR) is $36$~dB, the rate achieved by both users in MIMO-NOMA under fair power allocation is $4$~BPCU. This represents an expressive reduction of almost $3$~BPCU to the strong user's rate. On the other hand, the IRS-NOMA, in which we consider fixed reflection coefficients, can provide a high data rate performance for all users. The rate curves become superior even to that achieved by the strong one with fixed allocation, for the majority of the considered SNR range. This confirms that IRS-NOMA can bring high-performance fair networks to reality in the future.

\begin{figure}[t]
	\centering
	\includegraphics[width=.5\textwidth]{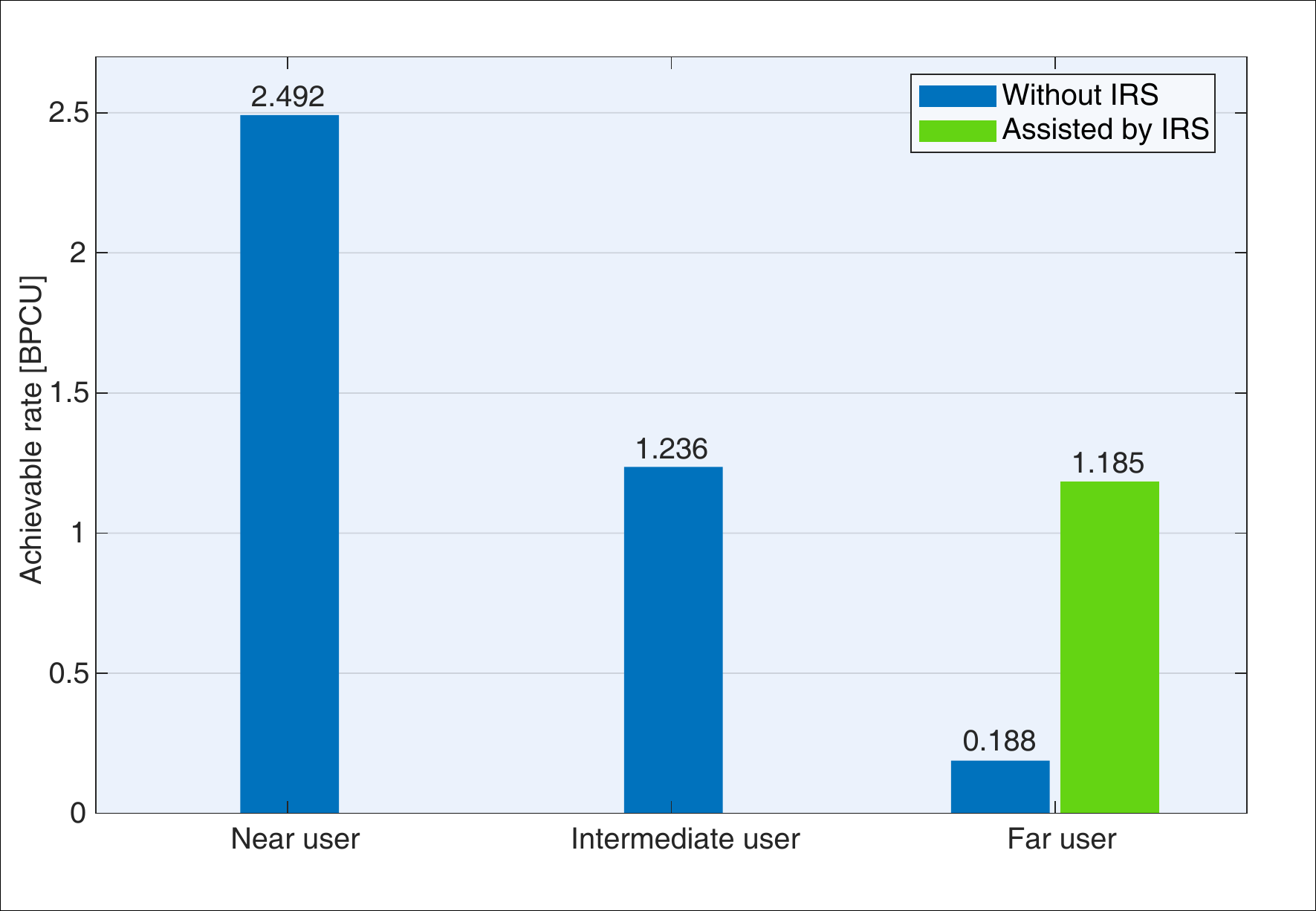}
	\caption{Achievable rates when the SNR is $26$~dBm for MIMO-NOMA and IRS-NOMA systems serving three users with fixed power allocation. The near user is located at $100$~m, the intermediate user at $200$~m, and the far user at $1500$~m from the BS. The power coefficients for the near, intermediate, and far users are $1/10$, $3/10$ and $6/10$, respectively.}\label{g3}
\end{figure}

\subsection{Enhanced Coverage Range}
In conventional MIMO-NOMA networks, it is difficult to provide uniform signal coverage to all existing devices. Users that are far from a BS, or that are suffering heavy blockage may experience poor or no signal reception. This issue becomes more relevant at the higher frequency bands of 5G, 6G, and beyond. The short wavelengths of the millimeter-wave and terahertz spectrum can resonate with atmospheric oxygen and water molecules, making a significant part of the radiated energy to be dissipated through kinetic absorption, causing strong signal attenuation. Such a harmful characteristic can be detrimental to the practical implementation of MIMO-NOMA. Specifically, serving users that are facing a too degraded signal reception with NOMA can lead to a substantial decrease in the system sum-rate. One could increase the power allocated to strong users to improve the sum-rate, but this strategy intensifies the interference at weak users, making their performance even worse. Consequently, users that are suffering from severe channel attenuation usually end up being disconnected, which limits the coverage range of practical MIMO-NOMA networks. Fortunately, IRSs are capable of enabling long-range communication to NOMA systems. As shown in Fig. \ref{achiev}c, an IRS could be installed close to users that are located in regions with no signal reception. As a result, more users are allowed to be served with NOMA, thereby, enhancing the connectivity capacity of such systems.

A practical demonstration of the aforementioned IRS capability can be seen in Fig. \ref{g3}. This simulation example shows the achievable rates for MIMO-NOMA and IRS-NOMA networks under fixed power allocation when the SNR is $26$~dBm. More specifically, we consider three users in the group of interest. Two users are closer to the BS, with the nearest user located at $100$~m and the intermediate one at $200$~m, while the third user is very far from the BS, located at $1500$~m. As one can observe, in the conventional MIMO-NOMA system, due to the long distance from the BS, the far user is experiencing a weak signal reception that allows a data rate of only $0.188$~BPCU. When the same user is assisted by an IRS, its rate is improved to $1.185$~BPCU, which represents a gain of more than 6 times when compared with that achieved in MIMO-NOMA and incredibly almost the same rate obtained by the intermediate user. This result clearly demonstrates another attractive improvement that can be achieved with IRS-NOMA systems.

\subsection{High Energy Efficiency}

Because NOMA exploits the power domain, the achievable rates of served users become highly coupled with their power allocation coefficients. For example, a direct way to improve the data rate of a weak user in a conventional MIMO-NOMA system can be achieved by increasing its power coefficient and decreasing the coefficients of other users. However, since too much power can be spent to achieve just a modest increase in the data-rate of one particular user, and since the system sum-rate is actually decreased, this strategy renders a low energy efficiency to MIMO-NOMA systems.

On the other hand, as it was demonstrated in the previous subsection, IRSs can improve the data rates of weak users without requiring more transmit power. Achieving this benefit and not increasing the energy consumption of the network is another potential advantage of IRS-NOMA systems. Specifically, if we consider a centralized IRS deployment as explained in Subsection \ref{iiib}, the only energy required will be to enable the IRS's reconfigurability capability, which can be implemented with the help of ultra-low-power electronics. In consequence, the use of energy harvesting components can be enough to supply all the necessary power, providing the IRS technology with an opportunity to become truly energy-neutral. As shown in Fig. \ref{achiev}d, if this attractive feature becomes a reality, we can achieve higher performance gains with less transmit power, significantly improving the energy efficiency of IRS-NOMA networks. Nevertheless, it is noteworthy that, if the energy neutrality assumption cannot be satisfied, the energy efficiency will inevitably be decreased.

\begin{figure}[t]
	\centering
	\includegraphics[width=.5\textwidth]{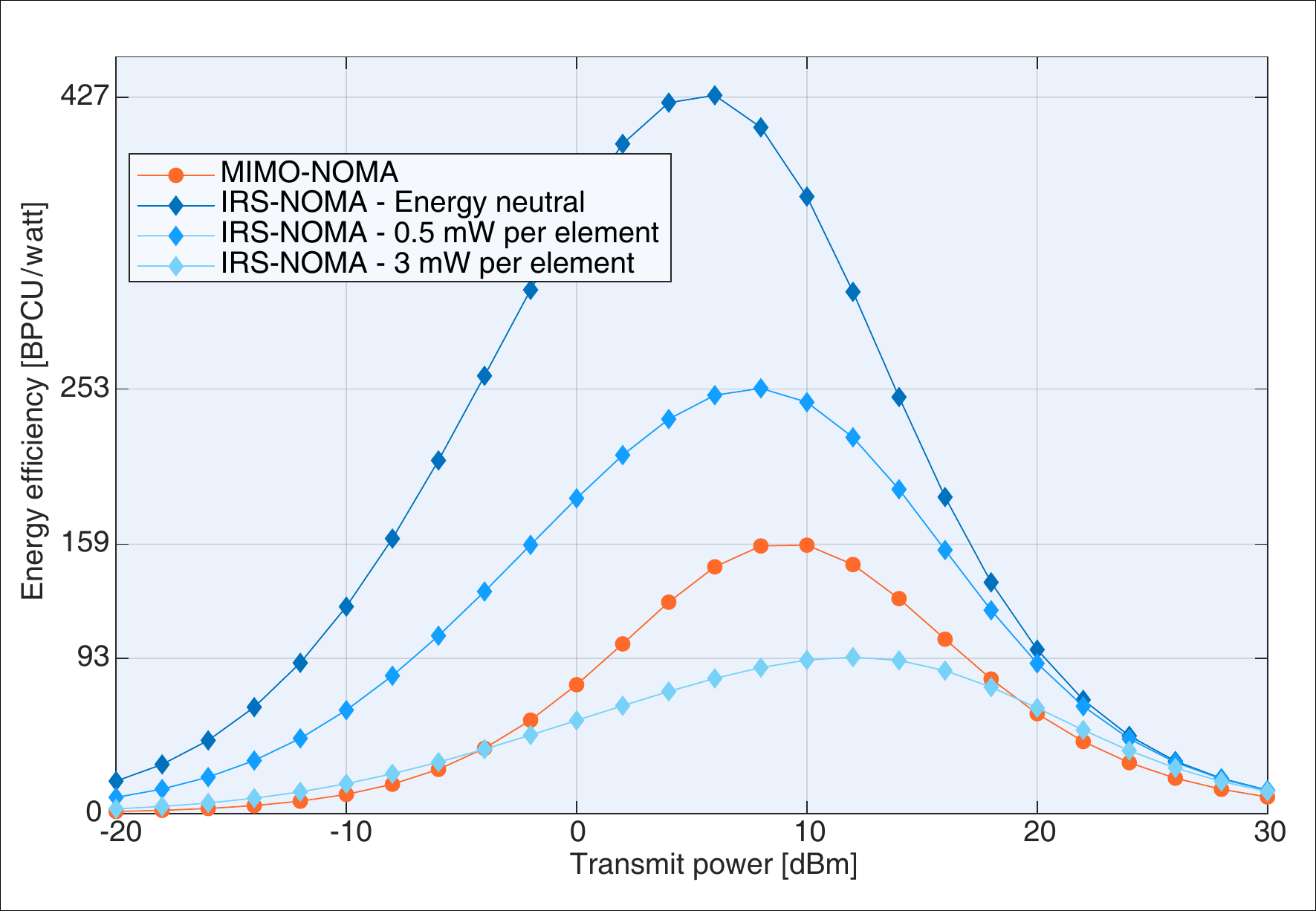}
	\caption{Energy efficiency versus transmit power for MIMO-NOMA and IRS-NOMA systems under fixed power allocation. Two users are considered in this simulation, one located at $100$~m and another at $200$~m from the BS. The power coefficients for the far and near users are $7/10$ and $3/10$, respectively.}\label{g4}
\end{figure}

In Fig. \ref{g4}, we present the energy efficiency curves versus transmit power for MIMO-NOMA and IRS-NOMA schemes. In order to show how energy efficient IRSs can become, we consider the scenario in which energy neutrality can be achieved, and as well as scenarios where the IRS contributes to the power consumption of the network. As one can see, when energy neutrality is considered, the IRS-NOMA system can offer remarkable energy efficiency improvements that outperform the conventional MIMO-NOMA counterpart. For example, when the transmit power is $10$~dBm, the conventional MIMO-NOMA system can reach a maximum energy efficiency of $159.6$~BPCU/Watt, while with the IRS-NOMA scheme, the maximum energy efficiency increases up to incredible $427.2$~BPCU/Watt and, at the same time, the transmit power required to reach this point decreases to $6$~dBm. However, when the IRS's energy consumption is taken into account, the energy efficiency is strongly impacted. For instance, if we consider that each reflecting element introduces an additional $0.5$~mW to the total power consumption, the maximum energy efficiency is decreased to approximately $253$~BPCU/Watt, and when the power consumption per element is $3$~mW, the energy efficiency becomes inferior to that achieved in the conventional MIMO-NOMA system.

\section{Scenarios and Opportunities}
In this section, we identify and discuss potential IRS-NOMA use case scenarios for future wireless networks.

\subsection{Multiple IRSs for Pervasive Coverage}\label{ssecmirs}

With the continuous growth of the global population, crowded environments are expected to become ever more common in the upcoming years. However, the majority of the existing IRS-NOMA related works consider system models where users are assisted by a single IRS, which are not suitable for these crowded scenarios of future networks. It can be extremely difficult, or even impractical, to optimize the meta-atoms of a single IRS to assist a large number of NOMA users with different channel gains and diverse requirements. Also, since users can be highly mobile, they might not always dispose of an IRS in range. To address this challenging use case scenario, one can envision a wide-scale deployment of multiple IRSs for pervasive coverage. In such scenarios, the IRSs could be jointly coordinated to deliver multiple independent beams to each user so that their channel gains could be flexibly tuned. This would enable the formation of small NOMA groups in crowded environments (a difficult task to accomplish relying solely on conventional MIMO-NOMA schemes), rendering a low SIC complexity to the users' devices. This large-scale IRS-NOMA network could provide massive access, ultra-high data rates, and ubiquitous signal coverage, enabling the deployment of futuristic applications such as holographic augmented reality and telepresence. For instance, in the future, large-scale IRS-NOMA can be deployed in crowded environments, like shopping malls, to attend the massive number of connections and provide ultra-high throughput to the users' sophisticated holographic enabled smartphones.

\subsection{3D Coverage in NOMA-UAV Networks}
The deployment of unmanned aerial vehicles (UAV) as aerial BSs is another appealing approach for improving signal coverage of the communication networks. The intelligent arrangement of multiple UAVs combined with their high operating altitude enables them to enhance the coverage area through efficient dynamic 3D beamforming. UAV networks can offer many advantages, such as flexibility in increasing the number of transmit antennas (with any array pattern), the ability to avoid obstacles (avoiding signal blockage), and more \cite{ref13}. In addition to that, if NOMA is applied to UAV networks, it becomes possible to serve multiple devices with a single 3D beam, which is attractive for enabling massive access in ultra-dense dynamic environments.

Despite the above advantages, it can be challenging to a NOMA-enabled UAV to serve multiple geographically dispersed user groups. Specifically, if the NOMA groups are separated far apart, providing a satisfactory performance to everyone can demand an excessive amount of energy. This can be a major impairment since a UAV dispose of only a limited power supply, i.e., an on-board battery. Therefore, strategies for reducing energy consumption in NOMA-UAV networks are crucial for their successful implementation. The employment of IRSs can be very effective for accomplishing this objective as well. For instance, instead of deploying all UAVs as active transmitters, we could implement hybrid networks where part of the UAVs work as active, and others operate as smart reflective devices equipped with only light-weight passive IRSs. A UAV assisted IRS-NOMA system would provide an extended signal coverage by exploiting a combination of passive and active 3D beamforming, enabling an enhanced communication performance even to NOMA groups far from active UAVs. By relaxing the need for a high transmit power such systems could present a low energy consumption (if IRS energy neutrality is satisfied), prolonging the UAVs' lifetime.

\subsection{Massive NOMA-Based Grant-free Transmissions}\label{gfree}
Grant-free protocols have arisen as efficient approaches for reducing the high signaling overhead faced in traditional cellular networks \cite{ref14}. In particular, NOMA-based grant-free transmissions allow multiple devices to transmit information in the uplink using the same spectrum, and without requiring grant to radio resources from the BS. Such schemes can efficiently tackle collision issues and reduce communication delays, making them ideal for enabling critical applications with ultra-low latency requirements. However, when the number of connected devices grows, the probability of achieving similar channel gains at the BS also increases. This characteristic can lead to poor multiuser detection performance. For example, in massive machine type communication applications like in a car factory, such impairment could lead to a failure in the production line. The installation of IRSs in this scenario can reduce the likelihood of happening this serious issue. IRS-NOMA-based grant-free transmissions could guarantee the BS to be able to always distinguish different devices so that the network can operate with a stable performance even when the number of connections becomes large.

\section{Research Challenges and Future Directions}

In Section \ref{secpi}, we have shown with a simple two-user example that the order of users in IRS-NOMA networks can be controlled by properly configuring the IRS's meta-atoms, providing more flexibility for optimizing the network performance. Despite the advantages of this capability, determining the optimal user ordering in a multi-user NOMA scenario can be challenging. This is because the IRSs must be dynamically optimized based on the instantaneous realizations of combined channels of direct and BS-IRS-user links, which are also dependent on the current set of meta-atoms. As a result, the users' effective channel gains (and, consequently, the user ordering) become coupled with the IRS coefficients. The resulting optimization problem becomes highly complex as the number of users and the number of reflecting elements increase, which imposes major challenges for the practical implementation of IRS-NOMA networks. Therefore, the development of efficient low-complexity algorithms for jointly optimizing the IRS coefficients and user ordering is mandatory and, thereby, a promising research direction.

User clustering schemes also play an important role in the performance of IRS-NOMA systems. For instance, in \cite{ref9}, it was shown that if users with certain channel conditions are grouped, IRS-NOMA can be outperformed by the IRS-OMA counterpart. This demonstrates that the design of efficient user clustering algorithms is also essential to exploit at maximum the benefits of IRS-NOMA systems. Given that only a few works have investigated this subject until now, this field of research allows for excellent opportunities of future studies. Furthermore, the fact that IRS-OMA can outperform IRS-NOMA also raises another issue: when should we use NOMA, and when should we use OMA? Such a fundamental question is not yet totally clarified and deserves further investigation.

Most of the contributions of this article were focused on the downlink. However, as demonstrated in Subsection \ref{gfree} with the practical example of NOMA-based grant-free transmissions, IRSs can also find useful applications in uplink MIMO-NOMA scenarios. For example, IRSs could be installed close to the users to amplify uplink transmissions, and as well as at the BS to ensure that the channel gains of the different users are always distinct. One could also deploy IRSs at BSs in shield mode to mitigate inter-cell interference in multi-cell uplink NOMA scenarios. Nevertheless, the challenges and trade-offs associated with these tempting use case scenarios are still unclear, and demand further studies.

The idea of a large-scale deployment of IRSs, as suggested in Section \ref{ssecmirs}, can also bring challenges to the network. Transmissions coming from one IRS can leak to others and cause strong inter-IRS interference, impacting the network performance. Such impairment could be mitigated if those IRSs with exceeding levels of interference could work in cooperative mode, as in \cite{refnew}, instead of operating independently. However, the joint coordination of multiple IRSs can trigger an explosion in signaling and processing overhead, which can potentially impact communication latency. In addition, since the channel gains observed at the receivers will be the result of the combination of all links from all IRSs, achieving the optimal set of reflection coefficients for determining the best user ordering for the entire IRS-NOMA network can become extremely difficult. Nevertheless, this is an interesting worth studying topic that is still lacking in the literature.

Finally, the concept of hybrid IRS-NOMA UAV networks (with part of the UAVs active and part passive) is another subject that can find attractive applications in 5G and beyond,  but that also leads to new unsolved problems. For example, UAV-to-ground and UAV-to-UAV channels are highly dynamic and have specific features that differ from those of conventional systems \cite{ref15}. As a result, the channels of IRSs mounted on UAVs will also exhibit unique characteristics that are still unknown. The accurate channel characterization of IRSs in the air, the development of specialized optimization frameworks, and the in-depth understanding of the application of NOMA in such scenarios arise then as promising research possibilities.



	\section*{Biographies}\vspace{-2mm}
	\vskip -1\baselineskip plus -1fil
	\begin{IEEEbiographynophoto}{Arthur Sousa de Sena}[M] (arthur.sena@lut.fi) received the B.Sc. degree in Computer Engineering and the M.Sc. degree in Teleinformatics Engineering from the Federal University of Ceará, Brazil, in 2017 and 2019, respectively. From 2014 to 2015, he studied Computer Engineering as an exchange student at Illinois Institute of Technology, USA. He is currently working toward the Ph.D. degree at the School of Energy Systems at LUT University, Finland. He is also a Researcher in the Cyber-Physical Systems Group at LUT. His research interests include signal processing, mobile communications systems, non-orthogonal multiple access techniques, intelligent metasurfaces, and massive MIMO.
	\end{IEEEbiographynophoto}

    \vskip -1\baselineskip plus -1fil
    
	\begin{IEEEbiographynophoto}{Dick Carrillo Melgarejo}[M] (dick.carrillo.melgarejo@lut.fi) received the M.Sc. degree in electrical engineering from Pontifical Catholic University of Rio de Janeiro, Rio de Janeiro, Brazil, in 2008. Between 2008 and 2010, he has worked as a research engineer on several research projects and standardization activities on cellular networks. From 2010 to 2018, he worked with the design and implementation of development projects based on LTE-Advanced, and LTE-Advanced Pro. Since 2018 he is a researcher at Lappeenranta--Lahti University of Technology, where he is also pursuing the Ph.D. degree in electrical engineering. His research interests are cellular networks beyond 5G supporting industrial verticals, and deep learning in communications.
	\end{IEEEbiographynophoto}
	
	\vskip -1\baselineskip plus -1fil

	\begin{IEEEbiographynophoto}{Fang Fang}[M] (fang.fang@durham.ac.uk) received her Ph.D. degree in electrical engineering from the University of British Columbia (UBC), Canada, in 2017. From 2018 to 2020, she was a Research Associate with the Department of Electrical and Electronic Engineering, The University of Manchester, UK.  Since August 2020, she has been with the Department of Engineering at Durham University, Durham (UK) as an Assistant Professor. Her current research interests include 5G and beyond wireless networks, NOMA, machine learning, and mobile edge computing.
	\end{IEEEbiographynophoto}
	
	\vskip -1\baselineskip plus -1fil

	\begin{IEEEbiographynophoto}{Pedro H. J. Nardelli}[SM] (pedro.nardelli@lut.fi)	is Assistant Professor (tenure track) and Academy Research Fellow at LUT University, Finland. He currently leads the Cyber-Physical Systems Group at LUT and is Project Coordinator of FIREMAN European consortium. He is also Adjunct Professor at University of Oulu. His research focuses on wireless communications particularly applied in industrial automation and energy systems.
	\end{IEEEbiographynophoto}
	
	\vskip -1\baselineskip plus -1fil

	\begin{IEEEbiographynophoto}{Daniel Benevides da Costa}[SM] (danielbcosta@ieee.org) was born in Fortaleza, Ceará, Brazil, in 1981. He received the B.Sc. degree in Telecommunications from the Military Institute of Engineering (IME), Rio de Janeiro, Brazil, in 2003, and the M.Sc. and Ph.D. degrees in Electrical Engineering, Area: Telecommunications, from the University of Campinas, SP, Brazil, in 2006 and 2008, respectively. His Ph.D thesis was awarded the Best Ph.D. Thesis in Electrical Engineering by the Brazilian Ministry of Education (CAPES) at the 2009 CAPES Thesis Contest. Since 2010, he has been with the Federal University of Ceará, where he is currently an Associate Professor.
	\end{IEEEbiographynophoto}
	
	\vskip -1\baselineskip plus -1fil
	
	\begin{IEEEbiographynophoto}{Ugo Silva Dias}[SM] (ugodias@ieee.org) received the B.Sc. degree in Electrical Engineering from the Federal University of Par\'{a}, Brazil, in 2004, and the M.Sc. and Ph.D. degrees in Electrical Engineering, from the State University of Campinas, Brazil, in 2006 and 2010, respectively. Since March 2010, Dr. Ugo Dias is an Assistant Professor at University of Brasilia (UnB), Brazil. He is a faculty member of the Department of Electrical Engineering. His main research interests include fading channels, field measurements, AI for future wireless networks, and wireless technologies in general. Prof. Dias is currently Editor of the \textsc{IET Electronics Letters} and \textsc{ACTA Press - Communications}.
	\end{IEEEbiographynophoto}
	
	\vskip -1\baselineskip plus -1fil

	\begin{IEEEbiographynophoto}{Zhiguo Ding}[F] (zhiguo.ding@manchester.ac.uk) is currently a Professor at the University of Manchester. From Sept. 2012 to Sept. 2020, he has also been an academic visitor in Princeton University. Dr Ding’ research interests are 5G networks, signal processing and statistical signal processing. He has been serving as an Editor for IEEE TCOM, IEEE TVT, and served as an editor for IEEE WCL and IEEE CL. He received the EU Marie Curie Fellowship 2012-2014, IEEE TVT Top Editor 2017, 2018 IEEE COMSOC Heinrich Hertz Award, 2018 IEEE VTS Jack Neubauer Memorial Award, and 2018 IEEE SPS Best Signal Processing Letter Award.
	\end{IEEEbiographynophoto}
	
	\vskip -1\baselineskip plus -1fil

	\begin{IEEEbiographynophoto}{Constantinos B. Papadias}[F] (cpapadias@acg.edu) is the executive director of the Research, Technology and Innovation Network (RTIN) at the American College of Greece (ACG). He is also a Professor of Information Technology at ACG’s Deree College and Alba Graduate Business School and Adjunct Professor at Aalborg University and at the University of Cyprus. Prior to these, he was a researcher at Institut Eurécom (1992-1995), Stanford University (1995-1997) and Lucent Bell Labs (1997-2006), where he also served as Technical Manager (2001-2006). He was Adjunct Professor at Columbia University (2004-2005) and Carnegie Mellon University (2006-2011) and Professor at Athens Information Technology (2006-2019), where he also served as Dean (2014-2019). He is Fellow of the European Alliance of Innovation since 2019.
	\end{IEEEbiographynophoto}
	
	\vskip -1\baselineskip plus -1fil

	\begin{IEEEbiographynophoto}{Walid Saad}[F] (walids@vt.edu) received his Ph.D degree from the University of Oslo in 2010. Currently,  he is a Professor at the Department of Electrical and Computer Engineering at Virginia Tech where he leads the Network sciEnce, Wireless, and Security (NEWS) laboratory. His  research interests include wireless networks, machine learning, game theory, cybersecurity, unmanned aerial vehicles, cellular networks, and cyber-physical systems. Dr. Saad was the author/co-author of eight conference best paper awards  and of the 2015 IEEE ComSoc Fred W. Ellersick Prize. He is an IEEE Fellow and an IEEE Distinguished Lecturer.
	\end{IEEEbiographynophoto}


\begin{thebibliography}{1}
	
	
	
	\bibitem{ref1} A. S. de Sena, D. B. da Costa, Z. Ding, and P. H. J. Nardelli, ``Massive MIMO-NOMA Networks with Multi-Polarized Antennas,'' {\it IEEE Trans. Wireless Commun.}, vol. 18, no. 12, Dec. 2019, pp. 5630-5642.
	
	\bibitem{ref2} A. S. de Sena, D. B. da Costa, Z. Ding, P. H. J. Nardelli, U. S. Dias, and C. B. Papadias, ``Massive MIMO-NOMA Networks with Successive Sub-Array Activation,'' {\it IEEE Trans. Wireless Commun.}, vol. 19, no. 3, Mar. 2020, pp. 1622-1635.
	
\bibitem{ref10} W. Saad, M. Bennis, and M. Chen, ``A Vision of 6G Wireless Systems: Applications, Trends, Technologies, and Open Research Problems,'' \emph{IEEE Network}, to appear, 2020.
	
	\bibitem{ref5} C. Huang, S. Hu, G. C. Alexandropoulos, A. Zappone, C. Yuen, R. Zhang, M. Di Renzo, and M. Debbah, ``Holographic MIMO surfaces for 6G wireless networks: Opportunities, challenges, and trends,'' [Online] {\it Available: https://arxiv.org/abs/1911.12296}, accessed on Jan. 2020.
	
		\bibitem{ref4} G. Yang, X. Xu, and Y.-C. Liang, ''Intelligent reflecting surface assisted non-orthogonal multiple access,'' [Online] {\it Available: http://arxiv.org/abs/1907.03133}, accessed on Jan. 2020.
	
	
	\bibitem{ref7} Q. Wu, and R. Zhang, ``Towards Smart and Reconfigurable Environment: Intelligent Reflecting Surface Aided Wireless Network,'' {\it IEEE Commun. Mag.}, 2019, pp. 1-8.
	
	\bibitem{ref11} M. Jung, W. Saad, M. Bennis, and C. S. Hong, ``On the Optimality of Reconfigurable Intelligent Surfaces (RISs): Passive Beamforming, Modulation, and Resource Allocation,'' [Online] {\it Available: https://arxiv.org/abs/1910.00968}, accessed on Jan. 2020.
	
	\bibitem{refnew1} C. Liaskos, S. Nie, A. Tsioliaridou, A. Pitsillides, S. Ioannidis, and I. Akyildiz, ``A New Wireless Communication Paradigm through Software-Controlled Metasurfaces,'' {\it IEEE Wireless Commun.}, vol. 56, no. 9, Sep. 2018, pp. 162-169.
	
	
	\bibitem{ref9} J. Zhu, Y. Huang, J. Wang, K. Navaie, and Z. Ding, ``Power Efficient IRS-Assisted NOMA,'' [Online] {\it Available: https://arxiv.org/abs/1912.11768}, accessed on Jan. 2020.
	
	
	\bibitem{RFID2009} N. V. Huynh, D. T. Hoang, X. Lu, D. Niyato, P. Wang, and D. I. Kim, ``Ambient Backscatter Communications: A Contemporary Survey,'' {\it IEEE Commun. Surv. Tutorials},	vol. 20, no. 4, May. 2018, pp. 2889-2922.
	
	
	
		\bibitem{ref12} A. S. de Sena, F. R. M. Lima, D. B. da Costa, Z. Ding, P. H. J. Nardelli, U. S. Dias, and C. B. Papadias, ``Massive MIMO-NOMA Networks with Imperfect SIC: Design and Fairness Enhancement,'' {\it IEEE Trans. Wireless Commun.}, to appear, 2020.
	

	
	\bibitem{ref13} M. Mozaffari, W. Saad, M. Bennis, Y. Nam, and M. Debbah, ``A Tutorial on UAVs for Wireless Networks: Applications, Challenges, and Open Problems,'' {\it IEEE Commun. Surv. Tutorials}, 2019, pp. 2334 - 2360.
	
	\bibitem{ref14} Z. Ding, R. Schober, P. Fan, and H. V. Poor, ``Simple Semi-Grant-Free Transmission Strategies Assisted by Non-Orthogonal Multiple Access,'' {\it IEEE Trans. Commun.}, vol. 67, no. 6, Jun. 2019, pp. 4464 - 4478.
	
	\bibitem{refnew} Y. Han, S. Zhang, L. Duan and R. Zhang, "Cooperative Double-IRS Aided Communication: Beamforming Design and Power Scaling," {\it IEEE Wireless Commun. Lett.}, to appear, 2020.
	
	\bibitem{ref15} Q. Zhang, W. Saad, and M. Bennis, ``Reflections in the Sky: Millimeter Wave Communication with UAV-Carried Intelligent Reflectors'', IEEE Global Commun. Conf. (GLOBECOM), Waikoloa, HI, USA, Dec. 2019.
	
\end{thebibliography}
\end{document}